\documentclass[sigconf]{acmart}
\setcopyright{none}              
\acmConference[]{}{}{}   
\acmYear{}
\copyrightyear{}
\acmDOI{}
\acmISBN{}
\renewcommand\footnotetextcopyrightpermission[1]{} 

\AtBeginDocument{%
  }

\acmISBN{978-1-4503-XXXX-X/2018/06}

\begin{document}

\title{MakOne: Behavioural Data of University Students’ Smart Devices in Uganda}




\author{Michael Kizito}
\affiliation{%
  \institution{Makerere University}
  \city{Kampala}
  \country{Uganda}}
  \orcid{0000-0001-5661-9433}
\email{michael.kizito@mak.ac.ug}

\author{Ivan Kayongo}
\orcid{0009-0007-4429-7335}
\affiliation{%
  \institution{University of Trento}
  \city{Trento}
  \country{Italy}
}
\email{ivan.kayongo@unitn.it}

\author{Hawa Nyende}
\affiliation{%
  \institution{Makerere University}
  \city{Kampala}
  \country{Uganda}}
\email{hawa.nyende@mak.ac.ug}

\author{Halimu Chongomweru}
\affiliation{%
  \institution{Makerere University}
  \city{Kampala}
  \country{Uganda}}
\email{haleem.chongomweru@mak.ac.ug}

\author{Lillian Muyama}
\affiliation{%
  \institution{Makerere University}
  \city{Kampala}
  \country{Uganda}}
\email{lillian.muyama@mak.ac.ug}

\author{Roy Alia Asiku}
\affiliation{%
  \institution{University of Trento}
  \city{Trento}
  \country{Italy}}
\email{royalia.asiku@unitn.it}

\author{Alice Mugisha}
\affiliation{%
  \institution{Makerere University}
  \city{Kampala}
  \country{Uganda}}
\email{alice.mugisha@mak.ac.ug}

\renewcommand{\shortauthors}{Michael Kizito et al.}


\begin{abstract}
  Understanding student behaviour in higher education is essential for improving academic performance, supporting mental well-being, and informing institutional policies. However, most existing behavioural datasets originate from Western institutions and overlook the unique socioeconomic and infrastructural contexts of African institutions, limiting the global applicability of resulting insights. This paper introduces MakOne, a novel multimodal dataset collected over six weeks from 72 students at \textit{Makerere University, Kampala}, using iLog, a mobile sensing application. The dataset integrates passive smartphone sensor data—including location, physical activity, and screen usage—with ecological momentary assessments (EMAs) that capture students’ moods and daily routines. Designed to reflect the lived experiences of students in an African setting, MakOne offers a foundation for research in behaviour modeling, inclusive context-aware system design, mental health analytics, and culturally grounded educational technologies. It contributes a critical African perspective to the growing body of data-driven studies on student behaviour.
\end{abstract}



\maketitle

\section{Introduction}
\label{introduction}


Recent advances in mobile sensing and ecological momentary assessment (EMA) techniques have enabled the collection of fine-grained, real-time behavioural data in naturalistic settings, offering researchers a more nuanced understanding of how students interact with their surroundings and respond to academic and social demands \cite{harari2016using, wang2014studentlife}, yet their application in African higher education remains under explored.

Despite growing interest in mobile sensing for behavioural and mental health research, the majority of publicly accessible datasets come from high-income nations and frequently represent Western, urban, and technologically savvy student populations \cite{saeb2016relationship, ranjan2019radar}. As a result, there is a significant gap in the availability of contextually grounded datasets from African societies, where socio-economic, infrastructural, and cultural dynamics may have a distinct impact on student behaviour. 

The development of effective and equitable AI systems increasingly depends on the use of data that is contextually grounded. Studies have shown that models trained on data from one region often perform poorly when applied elsewhere, revealing critical limitations in generalization. For example, Meegahapola et al. \cite{meegahapola2023generalization} observed that mood inference models based on mobile sensing data varied significantly in performance across eight countries, with personalized models trained on locally collected data outperforming global or cross-country models. This highlights the need to collect data that reflects the specific behaviours, environments, and cultural contexts of the target population. Echoing this, Moorosi \cite{moorosiBetterDataSets2024} emphasizes that AI for Africa must be developed within Africa, using data that captures local realities, and values. Together, these insights underscore the necessity of collecting a local, African dataset—not only to improve model accuracy, but to ensure that the resulting technologies are relevant, fair, and empowering to the communities they aim to serve. Thus by adopting the approach and methodology described in \cite{diversityOne2025Busso}, we present a novel multi-modal dataset collected from university students at \textit{Makerere University, Kampala}, using iLog, a mobile-based system that combines passive sensor logging with context-triggered self-report surveys \cite{zeni2014multi, kayongo2025methodology}.

The dataset spans six weeks and includes anonymized data on physical activity (from accelerometers), device usage (e.g., screen on/off, app usage patterns), as well as time-stamped ecological momentary assessments of mood, and daily activities. All data were collected in situ using students' personal smartphones running the iLog application \cite{zeni2014multi, kayongo2025methodology}, which is designed for lightweight, low-burden data collection in low-resource settings. The combination of sensor signals and self-reports allows for a multi-dimensional view of student life, capturing how behaviours, emotional states, and environmental factors fluctuate over time and in response to academic routines.

The main contributions of this work are:
\begin{itemize} 
\item \textbf{A novel multimodal dataset on student behaviour in Africa.} We introduce MakOne, the first dataset of its kind to capture behavioural, environmental, and emotional states of university students in a Sub-Saharan African setting. It combines passive smartphone sensor data (e.g., GPS, accelerometer, screen state) with culturally adapted self-reports (e.g., mood, activity, location). Unlike prior datasets from high-income contexts \cite{wang2014studentlife, vaizman2017recognizing, busso2025diversityone}, MakOne addresses a key geographic and demographic gap in ubiquitous computing. It enables cross-cultural comparisons, global model validation, and the development of more inclusive ubiquitous systems.
\item \textbf{A resource for cross-disciplinary ubiquitous computing research.} The dataset supports research in ubiquitous computing, human-computer interaction, and behavioural science. It enables applications ranging from machine learning (e.g., behaviour prediction) to digital wellbeing studies. As a high-quality dataset from an African university context, it helps researchers identify biases, test robustness across populations, and improve the inclusivity of ubiquitous systems. The dataset is a timely contribution for research communities interested in contextual awareness, mobile health (mHealth), and personalized interventions, particularly in the Global South.
\end{itemize}
\section{Related work}
\label{related work}
In this section we highlight related work that has been done in
Africa. Several studies have explored aspects of student life and behavior in African university settings, primarily using conventional survey-based methods. For example, Bantjes et al. \cite{bantjes2023treatment} conducted a large-scale self-report study across 17 South African universities to assess the prevalence of common mental disorders among students. Their findings underscored disparities affecting female, gender-nonconforming, and sexual minority students, especially those attending historically white institutions.

Ajiboye and Tella \cite{ajiboye2007university} examined the information-seeking behaviour of undergraduate students at the University of Botswana. Through structured questionnaires administered to 2,000 respondents, they found that academic information was the most sought-after, and that gender, level, and field of study influenced students’ information behaviour. Furthermore, Maphosa et al. \cite{maphosa2023student} used institutional data from the University of Johannesburg to investigate performance and dropout trends among engineering students, revealing gender disparities in enrollment and course satisfaction.

While these studies have generated valuable insights, they are methodologically limited in several key ways. Firstly, they rely exclusively on self-administered questionnaires or institutional records, which are subject to recall bias and limited temporal resolution. Secondly, they often lack sensor-based behavioural data, which restricts the granularity and ecological validity of behavioural modeling. Thirdly, and most importantly, they do not explicitly address diversity in data along multiple dimensions.

In contrast, the MakOne dataset builds on the DiversityOne framework \cite{diversityOne2025Busso}, whose dataset is available at \url{https://datascientia.disi.unitn.it/projects/diversityone}, which operationalizes diversity across three key axes: (i) sensor modality—capturing data from a rich set of smartphone sensors (e.g., accelerometer, GPS, screen, Bluetooth); (ii) participant diversity—capturing heterogeneity across demographics such as age, gender, and academic backgrounds; and (iii) contextual and cultural alignment—adapting self-report instruments to reflect local norms, and lifestyle patterns. MakOne is one of the first datasets in the African context to adopt this framework, integrating passive sensing and ecological momentary assessments (EMAs) through the iLog app, adapted specifically to local realities such as infrastructural limitations (e.g., intermittent internet and battery constraints).

By embracing these dimensions of diversity, MakOne fills a critical methodological gap in African behavioural research and contributes a valuable resource for cross-cultural comparisons, real-world behaviour modeling, and the development of equitable and context-aware ubiquitous systems.

\section{Methodology}
\label{methodology}
    The design and collection of the MakOne dataset focused on university students, leveraging their diversity to capture a multi-dimensional view of student life. The dataset reflects how behaviours, emotional states, and environmental factors fluctuate over time and in response to academic routines. To support this, the study combined structured self-report questionnaires with an intensive longitudinal study approach \cite{ranjan2019radar, saeb2016relationship}, enabled through an adapted version of the iLog app \cite{zeni2014multi, kayongo2025methodology}

    Drawing on the methodological framework of DiversityOne \cite{diversityOne2025Busso}, we operationalized diversity along three key dimensions: sensor modality, participant characteristics, and cultural-contextual alignment. While our data collection was limited to a single university in one country, we employed the same array of smartphone sensor modalities (e.g., accelerometer, GPS, screen status, Bluetooth) to ensure methodological consistency with established mobile sensing techniques. To improve cultural validity and ensure the relevance and interpretability of the self-report measures, we adapted the original questionnaires to reflect local norms and context-specific behaviours. For example, items related to modes of transport, types of food, and accommodation settings were revised to align with the lived experiences of our participants. Within the university setting, we also looked for heterogeneity across age, gender, and academic disciplines (undergraduate, and Masters from different colleges), enabling the study of behavioural variability within a localized yet demographically diverse cohort.

    The study protocol was developed using an adaptive, participatory approach to account for local perspectives and student needs. The original survey instruments and procedures were designed by a multidisciplinary team in the KnowDive group, at the University of Trento, Italy, comprising computer scientists, social scientists, interaction designers, incentive designers, ethicists, and legal experts. As local partners, we collaborated with the University of Trento, Italy to adapt these materials to suit our institutional, cultural, and infrastructural context. The adapted protocol was reviewed and validated by both our research team and the authorized institutional body to ensure compliance with ethical standards, privacy regulations, and methodological consistency.

    The MakOne survey collected detailed information on student lifestyles using a hybrid data collection strategy modeled on best practices from DiversityOne \cite{diversityOne2025Busso}. This strategy integrated passive smartphone sensor data with structured self-report prompts, enabling a multifaceted and temporally rich understanding of everyday behaviour. Such an approach provides high-resolution, fine-grained behavioural labels, supporting the development of machine learning models capable of capturing the complexities of daily life.

    Recruitment was carried out via university-wide email campaigns and informational flyers posted on noticeboards across all the colleges at the university. Each flyer included a QR code linking to an online registration form. Students who expressed interest through the registration form were contacted through the email addresses they provided and received detailed instructions in advance of data collection. This process supported effective coordination and participant engagement. To accommodate environments with intermittent internet connectivity, all survey questions were embedded directly within the iLog app \cite{zeni2014multi, kayongo2025methodology} to ensure timely delivery and offline access.

    Data collection took place over six weeks, from March 18 to April 29, 2024. During the first two weeks, participants received scheduled self-report prompts: a morning question at 08:00, an evening question at 22:00, a snack-related question every two hours, and time diary entries every 30 minutes between 08:00 and 22:00. This resulted in a maximum of 38 questionnaire items per participant per day. The remaining four weeks were dedicated solely to passive sensor data collection, after which the study concluded.

\section{Validation}
\label{validation}

The MakOne dataset encompasses a rich combination of self-reported and sensor-derived data, including
responses to questionnaires covering demographics, dietary habits,
activity levels, and modes of transportation among others, offering comprehensive insights into student behaviours within an African university context. This dual approach enables comparative analyses with datasets collected in other cultural settings, such as those referenced in \cite{diversityOne2025Busso}, while supporting a detailed understanding of how lifestyle, environmental factors, and daily routines interact locally.

During the study period, 73 participants consented to data collection and actively engaged with the survey and allowed sensor data collection via the iLog app \cite{zeni2014multi, kayongo2025methodology}. Out of these, 72 participants submitted at least one response, and 85\% (\textit{n}=62) answered one or more survey questions. Among respondents, the average reported age was 23.5 years (SD = 2.8), although approximately half chose not to disclose this information. Gender identification was provided by 44\% of respondents, with a distribution of 56\% male and 44\% female.

The intensive self-report component, including time diaries and scheduled questionnaires, was administered during the first two weeks of data collection. In contrast, passive sensor data were collected continuously throughout the entire six-week period. This design allowed for a high-resolution snapshot of participant-reported activities and states while minimizing reporting fatigue, complemented by longer-term sensor monitoring to capture behavioural patterns over time.

\begin{figure}[h]
    \centering
    \includegraphics[width=0.8\linewidth]{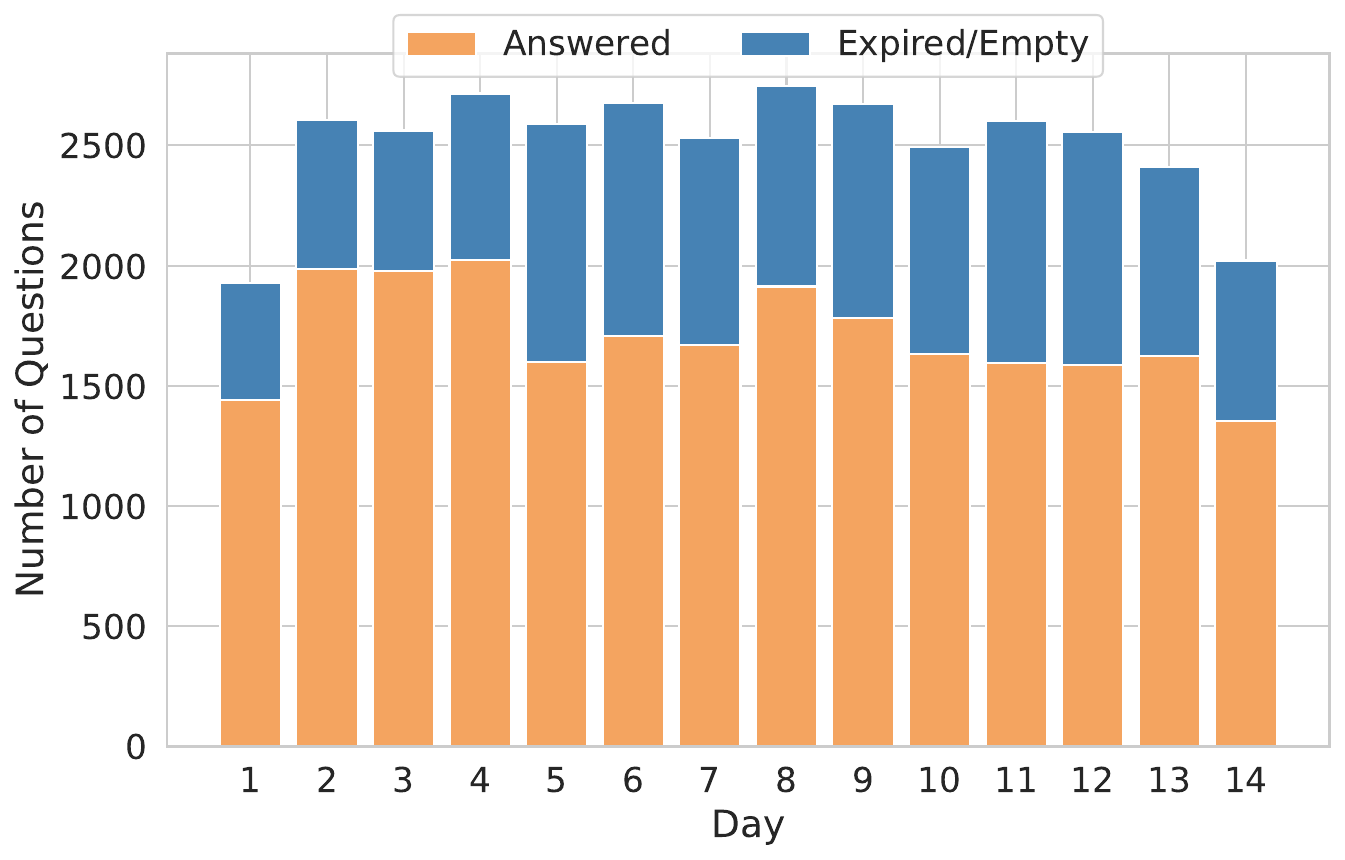}
    \caption{Total number of questions sent to participants per day. Orange: Questions with answers, Blue: questions without answers}
    \label{fig:daily_responses}
\end{figure}

\begin{figure}[h]
    \centering
    \includegraphics[width=0.7\linewidth]{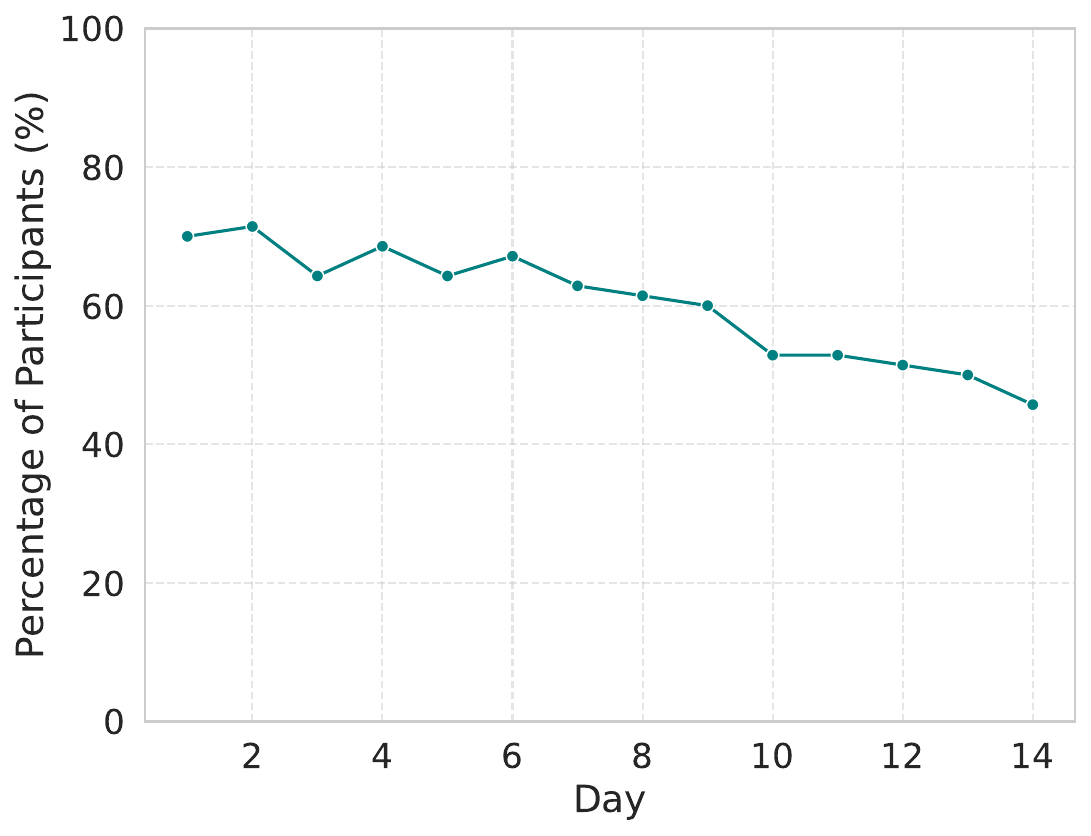}
    \caption{Percentage of participants that provided at least 20 responses per day.}
    \label{fig:active_participants}
\end{figure}

Engagement levels during the self-report phase were high, with approximately 65\% of participants submitting an average of 20 or more questionnaire responses daily. Figure~\ref{fig:daily_responses} illustrates the total number of questionnaire responses collected each day, while Figure~\ref{fig:active_participants} depicts the proportion of participants maintaining high response rates over time.

Sensor data collection yielded 20.94 GB of data from 30 smartphone sensors, categorized into seven domains: App Usage (applications, music, notifications), Connectivity (Bluetooth, cellular network, Wi-Fi), Device Usage (airplane mode, battery status, doze mode, headset events, ring mode, screen and touch events), Environment (pressure, light), Inertial (geomagnetic rotation vector), Motion (accelerometer, uncalibrated accelerometer, activities, gravity, gyroscope, rotation vector, step counter, step detector), and Position (location, magnetic field, orientation, proximity). Sensors such as the accelerometer collected continuous high-frequency data, whereas event-driven sensors like airplane mode logged data only upon state changes.

\begin{figure}[htbp]
  \centering
  \includegraphics[width=0.9\linewidth]{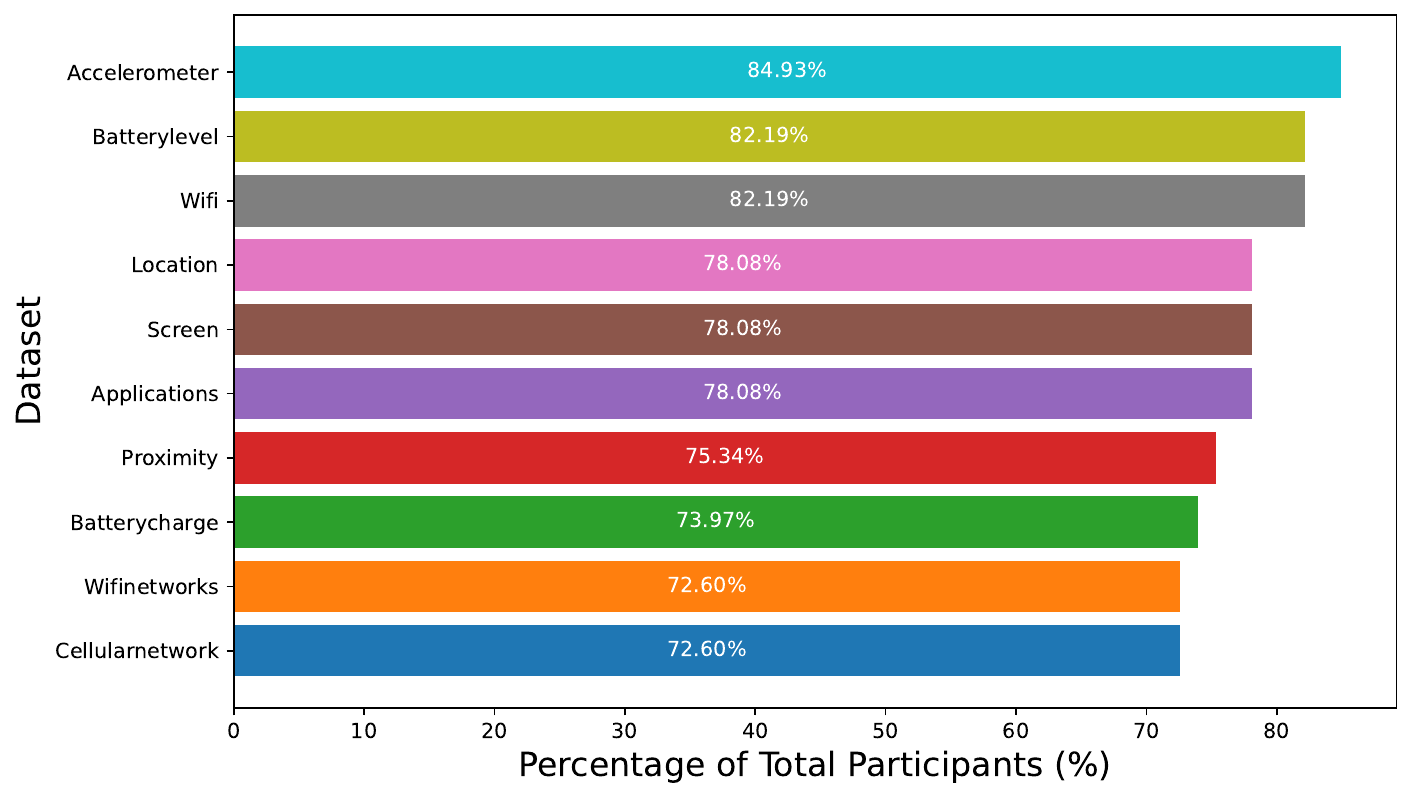}
  \caption{Top 10 sensor datasets and the proportion of unique participants who contributed data.}
  \label{fig:top_10_sensor_unique_participants}
\end{figure}

More than 80\% of unique participants contributed sensor data throughout the study, with the highest participation rates for accelerometer, Wi-Fi, and battery level sensors. In contrast, sensors such as gravity and activity recognition recorded data from fewer than 10\% of participants. Figure~\ref{fig:top_10_sensor_unique_participants} highlights the top ten sensor datasets by unique participant contribution.

\begin{figure}[h]
    \centering
    \includegraphics[width=0.9\linewidth]{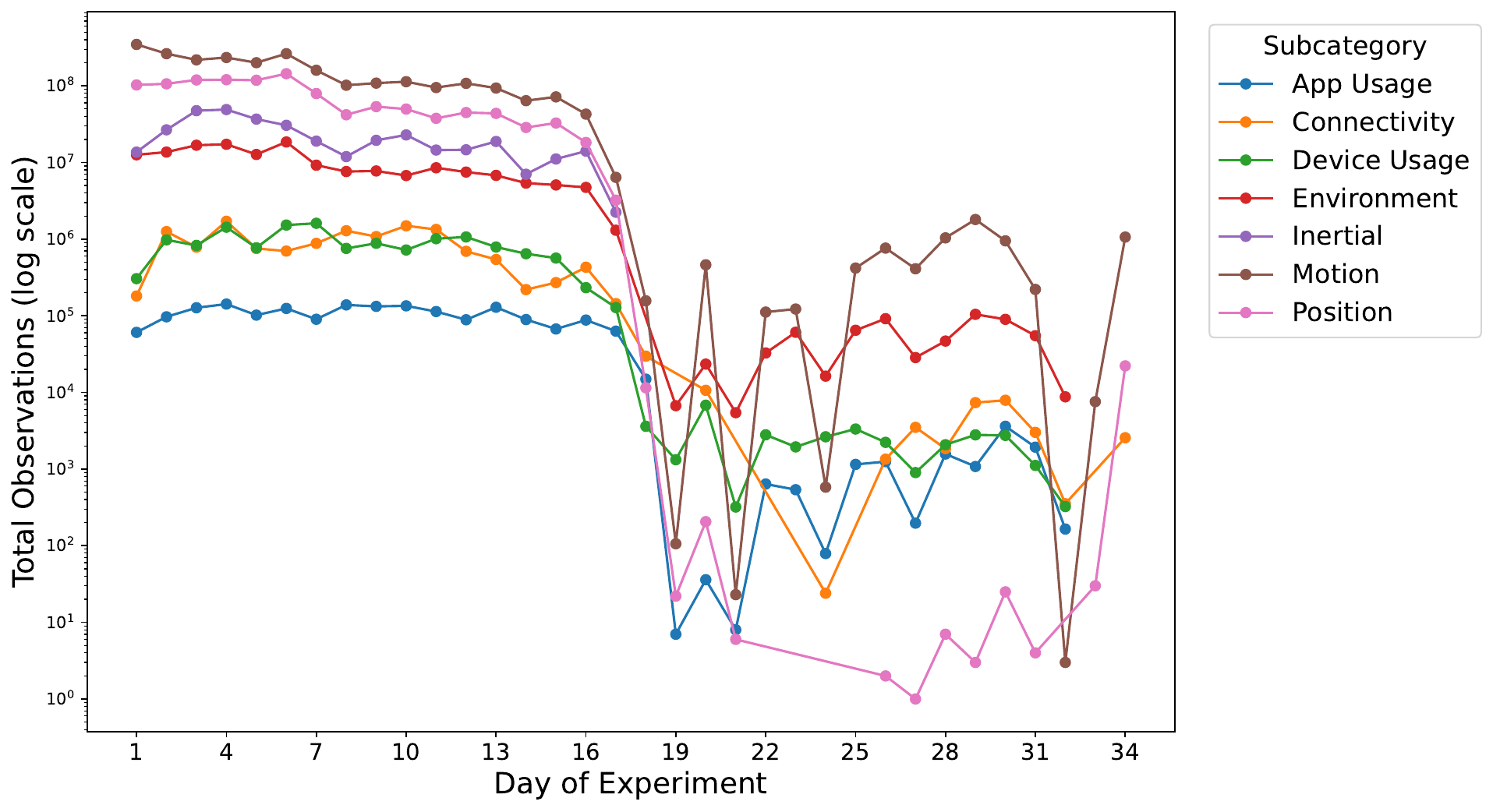}
    \caption{Daily total observations by sensor subcategory per day.} 
    \label{fig:daily_observations_subcategory}
\end{figure}

Participant compliance remained relatively stable throughout the study, as illustrated in Figure~\ref{fig:daily_observations_subcategory}, which displays the total number of daily sensor observations by subcategory on a logarithmic scale. Although a modest decline in data volume was observed over time, this trend indicates sustained participant engagement. Notably, a more pronounced drop occurred after day 17, likely due to some participants assuming the study had concluded when self-report questionnaires ended after the second week. Among the sensor subcategories, the App Usage category generated the fewest data points due to its event-driven logging (e.g., app openings or notifications), whereas the Motion category produced the largest volume of data, primarily from high-frequency, continuously sampled sensors such as the accelerometer.

Data quality analyses indicate minimal missing data in continuous sensor streams, with typical uptime exceeding 90\% for core sensors such as the accelerometer and GPS. Data gaps primarily resulted from participants disabling sensors or powering off devices, consistent with privacy choices and battery conservation.

Preliminary exploratory analyses demonstrate the dataset's potential to characterize daily activity patterns, identify transitions between academic and leisure periods, and model associations between environmental context and emotional self-reports. For example, combining accelerometer data with time diary entries facilitates classification of physical activity levels throughout the day, while location data supports context-aware behavioural modeling.

Compared to related datasets such as DiversityOne, MakOne contributes novel insights by focusing on an African university population and integrating culturally adapted self-report instruments. This enhances the dataset's relevance for studying behaviour in underrepresented contexts and enables valuable cross-cultural comparisons.
\section{Conclusion}
\label{conclusion}
This paper presents the MakOne dataset, a comprehensive multimodal resource capturing behavioural, environmental, and emotional data from university students within an African context. By integrating passive smartphone sensing with adaptive, culturally grounded self-report instruments, the dataset provides valuable opportunities for investigating complex daily life patterns among underrepresented populations. For example screen usage patterns can be examined across locations such as hostels, classrooms, or transit; and mood entries paired with social context and location reveal how emotional states shift depending on where students are and who they are with.

Throughout the study, several challenges emerged namely, ensuring consistent data quality was difficult due to the diversity in students’ living arrangements—ranging from campus halls to off-campus hostels and family homes—which affected both sensor and survey data collection. Additionally, intermittent internet connectivity and frequent phone battery depletion, especially when students were away from campus, impacted real-time data synchronization. The university’s diverse student body also exhibited varying behavioural routines; for example, while day students might attend classes during the day, evening students may be engaged in leisure activities, with their schedules reversed later. Moreover, intensive longitudinal studies of this kind are inherently susceptible to participant fatigue and dropout, which can reduce data completeness and introduce biases.

Despite these constraints, the MakOne dataset lays a robust foundation for cross-cultural behavioural modeling, global model validation, and the development of inclusive ubiquitous systems. Once publicly released, it will enable researchers to explore a broad range of topics across ubiquitous computing, human-computer interaction, behavioural psychology, and mobile health. Furthermore, by addressing ethical, contextual, and infrastructural nuances of data collection in an African setting, MakOne contributes to the advancement of responsible sensing and equitable representation in global data-driven research. Given that the dataset presents a slightly limited view to represent the diverse higher education scenarios in Africa, future work shall include conducting more pilot studies to collect more data to support the contribution of this paper.

\noindent
\textbf{Data availability} 
\newline
The dataset, along with detailed metadata, documentation, and usage guidelines, is available at:
\url{https://livepeople.datascientia.eu}.
Due to ethical and compliance considerations, access to the data is restricted and available only upon request. Interested researchers may submit a request by completing the form provided under the Catalogue section. We encourage prospective users to review the accompanying metadata and study protocol documentation to understand the dataset’s structure, variable definitions, and embedded privacy safeguards.

\begin{acks}
We thank colleagues from the KnowDive group, under the University of Trento for the help they provided in conducting this research.
\end{acks}

\bibliographystyle{ACM-Reference-Format}
\bibliography{main}

\begin{thebibliography}{10}

\bibitem{ajiboye2007university}
{\sc Ajiboye, J.~O., and Tella, A.}
\newblock University undergraduate students' information seeking behaviour: Implications for quality in higher education in africa.
\newblock {\em Turkish Online Journal of Educational Technology-TOJET 6}, 1 (2007), 40--52.

\bibitem{bantjes2023treatment}
{\sc Bantjes, J., Kessler, M.~J., Hunt, X., Stein, D.~J., and Kessler, R.~C.}
\newblock Treatment rates and barriers to mental health service utilisation among university students in south africa.
\newblock {\em International Journal of Mental Health Systems 17}, 1 (2023), 38.

\bibitem{diversityOne2025Busso}
{\sc Busso, M., Bontempelli, A., Malcotti, L.~J., Meegahapola, L., Kun, P., Diwakar, S., Nutakki, C., Britez, M. D.~R., Xu, H., Song, D., Correa, S.~R., Mendoza-Lara, A.-R., Gaskell, G., Stares, S., Bidoglia, M., Ganbold, A., Chagnaa, A., Cernuzzi, L., Hume, A., Chenu-Abente, R., Asiku, R.~A., Kayongo, I., Gatica-Perez, D., de~G\"{o}tzen, A., Bison, I., and Giunchiglia, F.}
\newblock Diversityone: A multi-country smartphone sensor dataset for everyday life behavior modeling.
\newblock {\em Proc. ACM Interact. Mob. Wearable Ubiquitous Technol. 9}, 1 (Mar. 2025).

\bibitem{busso2025diversityone}
{\sc Busso, M., Bontempelli, A., Malcotti, L.~J., Meegahapola, L., Kun, P., Diwakar, S., Nutakki, C., Britez, M. D.~R., Xu, H., Song, D., et~al.}
\newblock Diversityone: A multi-country smartphone sensor dataset for everyday life behavior modeling.
\newblock {\em Proceedings of the ACM on Interactive, Mobile, Wearable and Ubiquitous Technologies 9}, 1 (2025), 1--49.

\bibitem{harari2016using}
{\sc Harari, G.~M., Lane, N.~D., Wang, R., Crosier, B.~S., Campbell, A.~T., and Gosling, S.~D.}
\newblock Using smartphones to collect behavioral data in psychological science: Opportunities, practical considerations, and challenges.
\newblock {\em Perspectives on Psychological Science 11}, 6 (2016), 838--854.

\bibitem{kayongo2025methodology}
{\sc Kayongo, I., Malcotti, L., Zhao, H., and Giunchiglia, F.}
\newblock A methodology and a platform for high-quality rich personal data collection\#.
\newblock {\em The European Journal on Artificial Intelligence\/} (2025), 30504554251333615.

\bibitem{maphosa2023student}
{\sc Maphosa, M., Doorsamy, W., and Paul, B.~S.}
\newblock Student performance patterns in engineering at the university of johannesburg: An exploratory data analysis.
\newblock {\em IEEE Access 11\/} (2023), 48977--48987.

\bibitem{meegahapola2023generalization}
{\sc Meegahapola, L., Droz, W., Kun, P., De~G{\"o}tzen, A., Nutakki, C., Diwakar, S., Correa, S.~R., Song, D., Xu, H., Bidoglia, M., et~al.}
\newblock Generalization and personalization of mobile sensing-based mood inference models: an analysis of college students in eight countries.
\newblock {\em Proceedings of the ACM on interactive, mobile, wearable and ubiquitous technologies 6}, 4 (2023), 1--32.

\bibitem{moorosiBetterDataSets2024}
{\sc Moorosi, N.}
\newblock Better data sets won't solve the problem --- we need {{AI}} for {{Africa}} to be developed in {{Africa}}.
\newblock {\em Nature 636}, 8042 (Dec. 2024), 276--276.

\bibitem{ranjan2019radar}
{\sc Ranjan, Y., Rashid, Z., Stewart, C., Conde, P., Begale, M., Verbeeck, D., Boettcher, S., Dobson, R., Folarin, A., Consortium, R.-C., et~al.}
\newblock Radar-base: open source mobile health platform for collecting, monitoring, and analyzing data using sensors, wearables, and mobile devices.
\newblock {\em JMIR mHealth and uHealth 7}, 8 (2019), e11734.

\bibitem{saeb2016relationship}
{\sc Saeb, S., Lattie, E.~G., Schueller, S.~M., Kording, K.~P., and Mohr, D.~C.}
\newblock The relationship between mobile phone location sensor data and depressive symptom severity.
\newblock {\em PeerJ 4\/} (2016), e2537.

\bibitem{vaizman2017recognizing}
{\sc Vaizman, Y., Ellis, K., and Lanckriet, G.}
\newblock Recognizing detailed human context in the wild from smartphones and smartwatches.
\newblock {\em IEEE pervasive computing 16}, 4 (2017), 62--74.

\bibitem{wang2014studentlife}
{\sc Wang, R., Chen, F., Chen, Z., Li, T., Harari, G., Tignor, S., Zhou, X., Ben-Zeev, D., and Campbell, A.~T.}
\newblock Studentlife: assessing mental health, academic performance and behavioral trends of college students using smartphones.
\newblock In {\em Proceedings of the 2014 ACM international joint conference on pervasive and ubiquitous computing\/} (2014), pp.~3--14.

\bibitem{zeni2014multi}
{\sc Zeni, M., Zaihrayeu, I., and Giunchiglia, F.}
\newblock Multi-device activity logging.
\newblock In {\em Proceedings of the 2014 ACM international joint conference on pervasive and ubiquitous computing: adjunct publication\/} (2014), pp.~299--302.

\end{thebibliography}

\appendix

\end{document}